\documentclass[%
 reprint,
 amsmath,amssymb,
 aps,
]{revtex4-2}

\usepackage{graphicx}% Include figure files
\usepackage{dcolumn}% Align table columns on decimal point
\usepackage{bm}% bold math
\usepackage{bbm}

\begin{document}

\preprint{APS/123-QED}

\title{Phyllotaxis in a Keller-Segel model}

\author{Michael F. Staddon}
\affiliation{Max Planck Institute of Molecular Cell Biology and Genetics, Dresden, Germany}
\affiliation{Center for Systems Biology Dresden, Dresden, Germany}
\affiliation{Cluster of Excellence, Physics of Life, TU Dresden, Dresden, Germany}

% \author{Boris Shraiman}
% \affiliation{
%  Kavli Institute for Theoretical Physics, University of California, Santa Barbara, United States
%  }


\begin{abstract}

% TO DO: Need to discuss more from the review
% - Packing models
% - D&C like models
% - Plant biophysics, SAM, auxin source at tip and growth
% - Auxin-stress feedback model
% - Auxin-pins model

% Introduction
Plants often exhibit regular arrangements of seeds or leaves, known as phyllotaxis, and can famously exhibit Fibonacci spirals, as in sunflower heads or on pine cones. While there are many models which can reproduce spiral formation, the actual mechanism of arrangement remains unclear. Here, we test a more general class of model, based on the Keller-Segel model, in which a diffusing chemical signal produces its own transport signal and can form local instabilities. By modelling the plant as a disc  with a signal source at the center and radial growth, we show that we can reproduce both Fibonacci spirals and alternating patterns, depending on a balance between growth and source. Finally, using linear stability analysis, we show how the pattern arises due to a balance between the growth of instabilities and the growth of the plant. Overall, this work demonstrates how the Keller-Segel model can reproduce a wide range of observed patterns, and may act as a phenomenological description of a number of processes in plants.

\end{abstract}

\maketitle

%\tableofcontents

\section{Introduction}

The arrangement of leaves and branches in plants, known as phyllotaxis, can result in a variety of patterns~\cite{levitov1991fibonacci, adler1997history}. The most famous are the Fibonacci spirals, where each leaf has an approximately golden angle $\Phi = 180 (3-\sqrt{5})^\circ\approx137.5^\circ$ difference from the previous. When counting the number of spirals connecting nearest neighbours, one finds a pair of Fibonacci numbers, such as 34 spirals one way and 55 the other on sunflower heads, or 8 and 13 for pine cones. Alternatively, leaves may be whorled, where several leaves form at the same level along the stem and can form alternating pairs, as in the basil plant, or in groups of three leaves separated by $120^\circ$. Three main theoretical approaches have been used to model the phenomenon: geometric, mechanical, and biochemical, though the exact way leaves are arranged remains unclear~\cite{adler1997history, pennybacker2015phyllotaxis}.

% Geometric models
The earliest approaches to modelling phyllotaxis consisted of geometric studies following Hofmeister's model, which suggest that new leaves form at the least crowded part of the stem~\cite{hofmeister1868allgemeine}. Airy tested this theory by gluing spheres onto a stretched elastic band, which, when allowed to relax, packed into a cylindrical Fibonacci-spiral~\cite{airy1873leaf}. Van Iterson developed this idea further by considering the packing of hard discs around a cylinder and showed that it has an optimal packing that forms a regular lattice, with the angle between successive discs being determined by the ratio of disc and cylinder sizes~\cite{van1907mathematische}.

% Physical models
Later models considered self-organisation due to mechanical interactions between new leaves. Instead of using hard discs, Douady and Couder~\cite{douady1992phyllotaxis} considered the interaction of repulsive particles. In their experiments, ferrofluid droplets were added to the center of a disc at regular intervals, and were radially advected from the center. Due to the repulsive interactions between nearby droplets, they found that the divergence angle between successive drops was $180^\circ$ when the advection speed was high  and converged to the golden angle as the advection speed was decreased. Theoretically, it was shown that any repulsive particles arranged on a cylinder should form Fibonacci spirals~\cite{levitov1991energetic}, as was demonstrated by the construction of a ``mechanical cactus'' consisting of magnetic rods attached to a column that could freely rotate~\cite{nisoli2010annealing}. Alternatively, one could consider mechanical interactions from growth of the plant. Since the primordia grow faster than the surrounding tissue, the surface may buckle~\cite{dumais2000new} and can theoretically give Fibonnaci spirals through mechanics alone~\cite{shipman2005polygonal}.

% Biochemical models
Recent models have been more biologically informed, by considering the role of the plant hormone auxin in specifying the location of new leaves~\cite{zhao2010auxin, vernoux2010auxin, pinon2013local}. Auxin undergoes active polar transport by the PIN1 proteins, with regions of high auxin concentration specifying the location of the primordia which grow into the new leaves or branches~\cite{reinhardt2003regulation}. By including positive feedback between auxin and PIN1 polarisation~\cite{smith2006plausible, newell2008phyllotaxis} either directly~\cite{jonsson2006auxin}, or indirectly through growth-driven stress~\cite{heisler2010alignment, pennybacker2013phyllotaxis} or some biochemical signal such as calcium ions~\cite{lamport2020phyllotaxis}, models can form stable phyllotactic patterns where PIN1 polarises towards regions of high auxin concentration. However, the exact mechanism of alignment remains unclear, in particular since the precursors of primordia appear much earlier than the differential growth becomes apparent.

Here we consider a simplified description of morphogen-driven self-organisation, in the form of the Keller-Segel model~\cite{keller1971model, herrero1996chemotactic, arumugam2021keller}. Originally describing chemotaxis in bacteria, we can reinterpret the model in terms of cell signalling with two counter diffusing signals. One signal represents auxin, which produces a transport signal such as mechanical stress or a chemical signal. Auxin is then transported up the gradient of the transport signal in a similar manner to PIN1 transport. Thus, we agnostically replace polarisation with local secretion and diffusion. Since it is long range and locally produced, it is the simplest model of the types considering auxin-PIN1 or stress-PIN1 alignment~\cite{jonsson2006auxin, pennybacker2013phyllotaxis}. Using this model, we are able to obtain a range of phyllotactic patterns including Fibonacci spirals or whorls, with the type determined by both growth speed and rate of auxin production. Using a simple linear stability analysis, we can understand how growth and auxin source work together to determine the pattern.

\section{Modelling plant morphogenesis}

% Needs more of a summary - why do we use this specific model, because it causes self attraction
The classical Keller-Segel model for chemotaxis~\cite{keller1971model} considers the concentration of bacteria which produces its own chemoattractant. The cells execute a run-and-tumble style random walk, with a bias towards regions of higher chemoattractant concentration. In the continuum limit, we write $a$ as the concentration of bacteria, and $c$ as the concentration of chemoattractant, which obey
\begin{equation}
    \dot{a} = D_a \nabla^2 a - \alpha \nabla \cdot (a \nabla c)
\end{equation}
\begin{equation}
    \dot{c} = D_c \nabla^2 c + \beta a - \gamma c.
\end{equation}
The bacteria produces the chemoattractant with rate $\beta$, which in turn degrades at rate $\gamma$. Both bacteria and the chemoattractant diffuse, with coefficients $D_a$ and $D_c$ respectively. However, the bacteria also swims up the concentration gradient of the chemoattractant which with speed $\alpha$. At high enough concentrations of bacteria, the positive feedback can outcompete the diffusion and lead to instabilities which grow over time, concentrating the bacteria into local maxima which even diverges in finite time~\cite{herrero1996chemotactic}.

In this work, we reinterpret the Keller-Segel model in terms of auxin $a$ and a transport signal $c$, which could represent mechanical stress~\cite{pennybacker2013phyllotaxis} or a biochemical signal~\cite{lamport2020phyllotaxis}. Auxin diffuses and is actively transported up the gradient of the transport signal, representing PIN1 polarisation. We use a saturating rate of transport to prevent divergences, and work in dimensionless units by rescaling time by the degradation rate $1/\gamma$, and space by the diffusion rate $ \gamma \sqrt{D_u}$. We model the surface of the plant as a disc of radius $r_d$. To model plant growth, we include a source of $a$ in the center, $S(r, t)$, and a radial velocity field representing plant growth, $\mathbf{v}$:
\begin{equation}
    \dot{a} + \nabla \cdot (a \mathbf{v}) = \nabla^2 a - \alpha \nabla \cdot \left(a \nabla \frac{c}{c + c_0}\right) + S
\end{equation}
\begin{equation}
    \dot{c} + \nabla \cdot (c \mathbf{v}) = D\nabla^2 c + \beta a - c
\end{equation}
where $\alpha$ controls the transport rate of auxin, $S$ is a source of auxin which potentially varies in space and time, $D$ is the diffusion coefficient for the transport signal $c$, and $\beta$ is the rate of the transport signal production by auxin. We use a saturating transport rate, with saturation parameter $c_0$, since the linear transport rate in the Keller-Segel model leads to singularities~\cite{herrero1996chemotactic}. While this represents a phenomenological model, we will call the field $a$ auxin, since peaks of auxin indicate the location for new primordia to grow, and will call $c$ the transport signal.

\begin{figure}[h]
\includegraphics[width=\columnwidth]{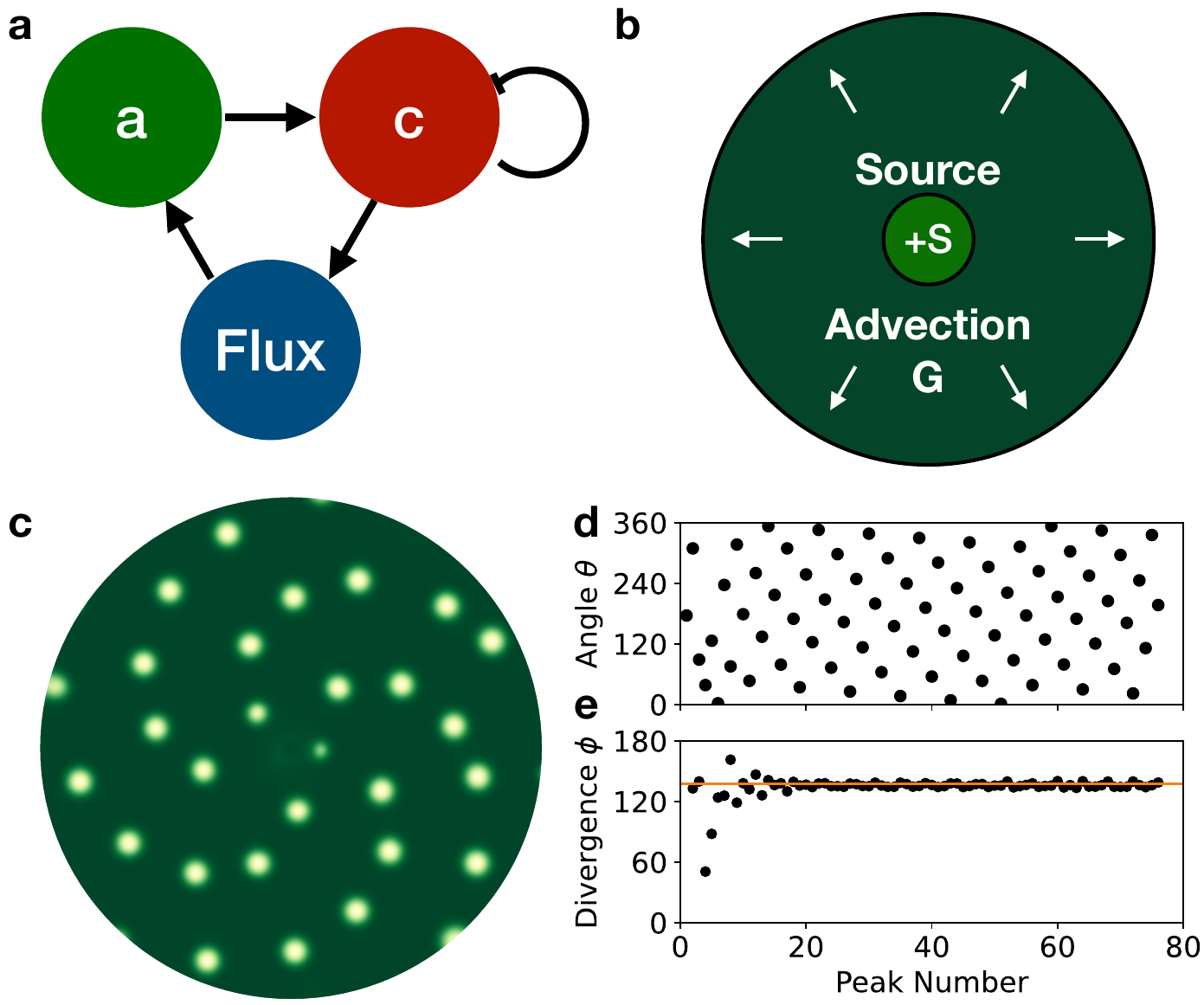}
\caption{A Keller-Segel model of plant morphogenesis produces spirals. (a) Schematic of the saturating Keller-Segel model. (b) Schematic of the system geometry. A source of auxin is added within the central region with a flux of $S$. A radial velocity advects material outward, controlled by the growth parameter $G$. (c) Simulation output with $S = 1$ and $G = 1.2$. (d) Polar angle of each auxin peak against peak number. (e) Divergence between successive peaks. The golden line indicates the golden angle $\Phi$.}
\label{fig:1}
\end{figure}

As in the Keller-Segel model for chemotaxis, positive feedback between auxin and its the transport signal through attractive fluxes can lead to pattern formation. In comparison to the ferromagnetic droplets used by Douady and Couder~\cite{douady1992phyllotaxis}, the peaks are repulsive at long ranges, as the two peaks compete for auxin, but are attractive at short range as they absorb nearby chemicals and merge instead of repelling.

We simulate the formation of patterns on a circular disc with radius $r_d = 50$, and initially zero $a$ and $c$ concentration. Within a central region of radius $r_s$, we add $a$ at rate $S$ per unit time and area. The source rate $S$ starts at zero flux but increases over time to its maximum rate $S_0$, which helps patterns to form more stably:
\begin{equation}
  S(r, t)=\begin{cases}
    S_0 (1 - e^{-t / \tau}), & \text{if $r<r_S$}.\\
    0, & \text{otherwise}.
  \end{cases}
\end{equation}
Material is then advected away from the center with velocity
\begin{equation}
    \mathbf{v} = \frac{G}{r} \hat{\mathbf{r}}
\end{equation}
representing growth of the plant, where $G$ is the area growth rate. The simulation is run until $t = 2000$, by which point a stable pattern has emerged, integrating the equations numerically using the finite volume package fipy~\cite{guyer2009fipy}.

% Needs improving
Using the parameters $S_0 = 1.0$ and $G = 1.2$, we integrate the equations over time and observe the evolution of the pattern. The source of $a$ increases the concentration in the system. At high enough concentration the system breaks symmetry and splits into several high concentration peaks which are advected away radially. New peaks are added sequentially between the gaps, eventually being added at a regular rate with a stable angle between successive peaks, or divergence. For these parameters, the mean divergence is $\bar{\phi} = 136^\circ$, close to the golden angle, $\Phi = 180 * (3 - \sqrt{5}) \approx 137^\circ$.

\begin{figure}[h]
\includegraphics[width=\columnwidth]{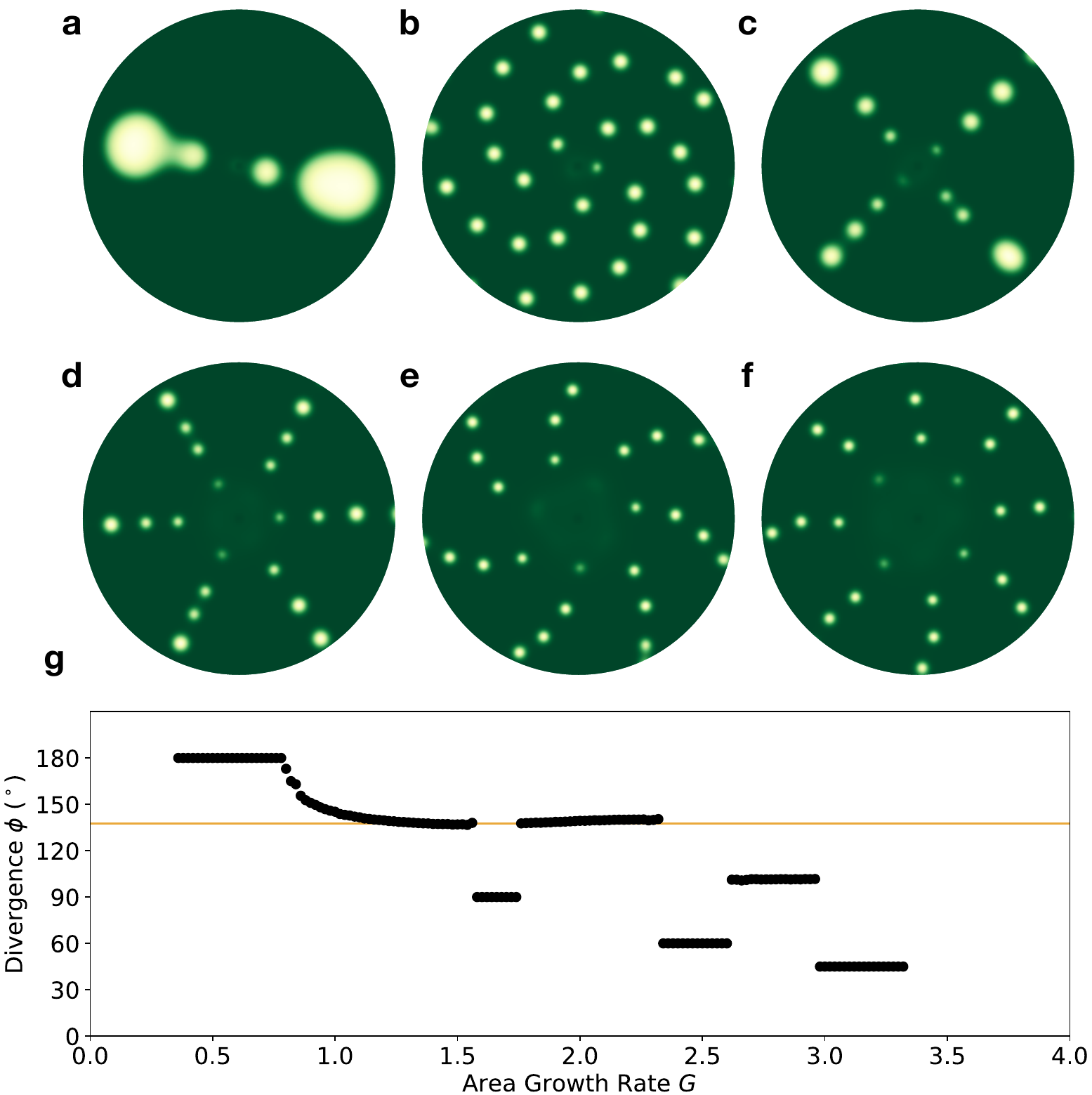}
\caption{The pattern depends on the growth rate. (a) A (1, 2) pattern, $S_0 = 1, G = 0.5$. (b) A Fibonacci spiral (2, 3) pattern, $S = 1, G = 1.2$. (c) An alternating (2, 4) pattern, $S_0 = 1, G = 1.6$. (d) A (3, 6) pattern, $S_0 = 1, G = 2.5$. (e) A (4, 7) pattern, $S_0 = 1, G = 2.9$. (f) A (4, 8) pattern, $S_0 = 1, G = 3.1$. (g) Divergence against area growth rate $G$. The golden line shows the golden angle, $\Phi \approx 137^\circ$.}
\label{fig:2}
\end{figure}

Different patterns are observed depending on the value of the growth rate, $G$ (Fig.~\ref{fig:2}g). For very low growth velocity, the auxin from the source is able to attract auxin faster than advection, trapping auxin in a large peak around the center of the disc. For low growth rate, around $G = 0.5$, auxin can escape the center and new peaks are added with a $180^\circ$ divergence, giving a (1, 2) pattern (Fig.~\ref{fig:2}a). Note that since peaks are attractive and the radial velocity decreases with distance, peaks eventually merge. While this is a problem in the model, within the plant the pattern is set within only a small distance from the center, after which peaks would stabilise and grow into primordia. As the velocity increases further to around $G = 0.8$, the peaks are added with a slight twist (Fig.~\ref{fig:2}b). Further increases to the growth rate decrease the divergence, which stabilises around the golden angle Fibonacci spiral (2, 3) patterns (Fig.~\ref{fig:2}c). At around $G = 1.5$ the Fibonacci spirals again become unstable, and instead we find alternating (2, 4) patterns, with pairs of peaks added at a time with a $90^\circ$ divergence (Fig.~\ref{fig:2}d), as seen in plants like basil. At around $G = 1.7$ Fibonacci spirals become stable. Continuing this trend, as the growth velocity keeps increasing the patterns successively increase in parastichy number, seeing (3, 6) patterns around $G = 2.4$, (4, 7) patterns around $G = 2.6$, and (4, 8) patterns around $G = 3.0$, beyond which the auxin is too dilute to form patterns and we obtain a uniform state.

Similarly, the rate of auxin addition at the center, $S_0$, can be varied. We find that increasing the auxin flux has the opposite effect to increasing the growth rate, with increases in $S_0$ leading to decreases in the paristichy number. For example, starting from $S_0 = 1$ and $G = 2$, the system has a Fibonacci (3, 5) pattern. Increasing $S_0$ eventually changes the pattern to an alternating (2, 4), while decreasing it can give (3, 6) and higher patterns before the concentration is too low for patterns to form. Such a trend exists for all values for $S_0$ and $G$, thus we next develop a simple model that predicts which patterns are expected using linear stability analysis.

\begin{figure}[h]
\includegraphics[width=\columnwidth]{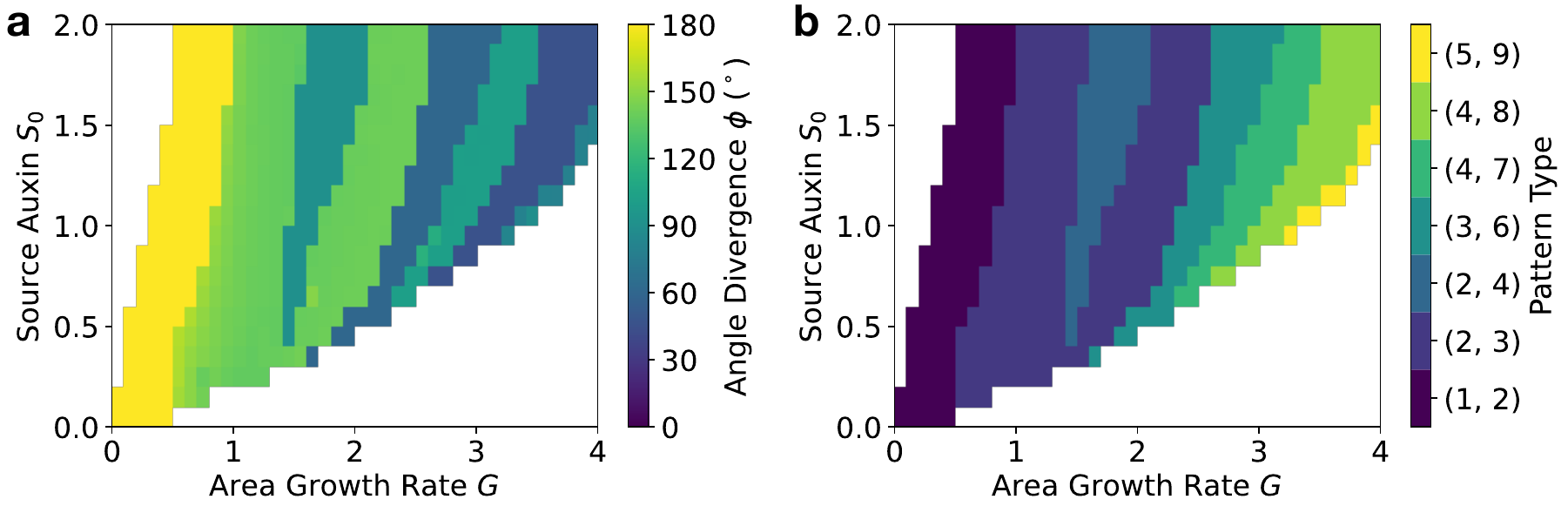}
\caption{Auxin flux and area growth rate control the pattern. (a) Angle divergence $\phi$, and (b) pattern type, against source auxin $S_0$ and area growth rate $G$. The unshaded regions indicate regions where patterns do not form, in the bottom right, or auxin cannot escape the center, in the top left.}
\label{fig:3}
\end{figure}

To understand these results, we construct a dimensionless number using the growth rate $G$ and the dominant wave number $k(G, S_0)$, and its corresponding growth rate $\lambda(G, S_0)$ from linear stability analysis (see Appendix), of the form
\begin{equation}
    Ro = \frac{G k^2}{\lambda}.
\end{equation}
which we call the Rosette number, as it aims to predict the arrangement of the peaks or leaves around the disc. Since $\lambda$ is the growth rate of the instability and has units of inverse dimensionless time, $1 / \lambda$ scales with the period between successive peaks, thus $G / \lambda$ represents the amount of area added to the shoot as the peaks form. Since $k$ is the wave number with units of inverse dimensionless length, $1/k$ scales with the distance between peaks, and so $1/k^2$ the area closest to each peak. Thus the Rosette number can be thought of as representing the relative spacing of the peaks, equal to the ideal area between successive peaks over the area added to the shoot from growth. When $Ro$ is high, then the peaks are relatively sparse, and we might expect alternating patterns, as new peaks only interact with the last added peak, as in the case of high growth rates in the Douady and Couder experiments~\cite{douady1992phyllotaxis}. When $Ro$ is low, then the peaks are relatively dense and we would expect new peaks to interact with many of the existing peaks, and so form complex spirals.

\begin{figure}[h]
\includegraphics[width=\columnwidth]{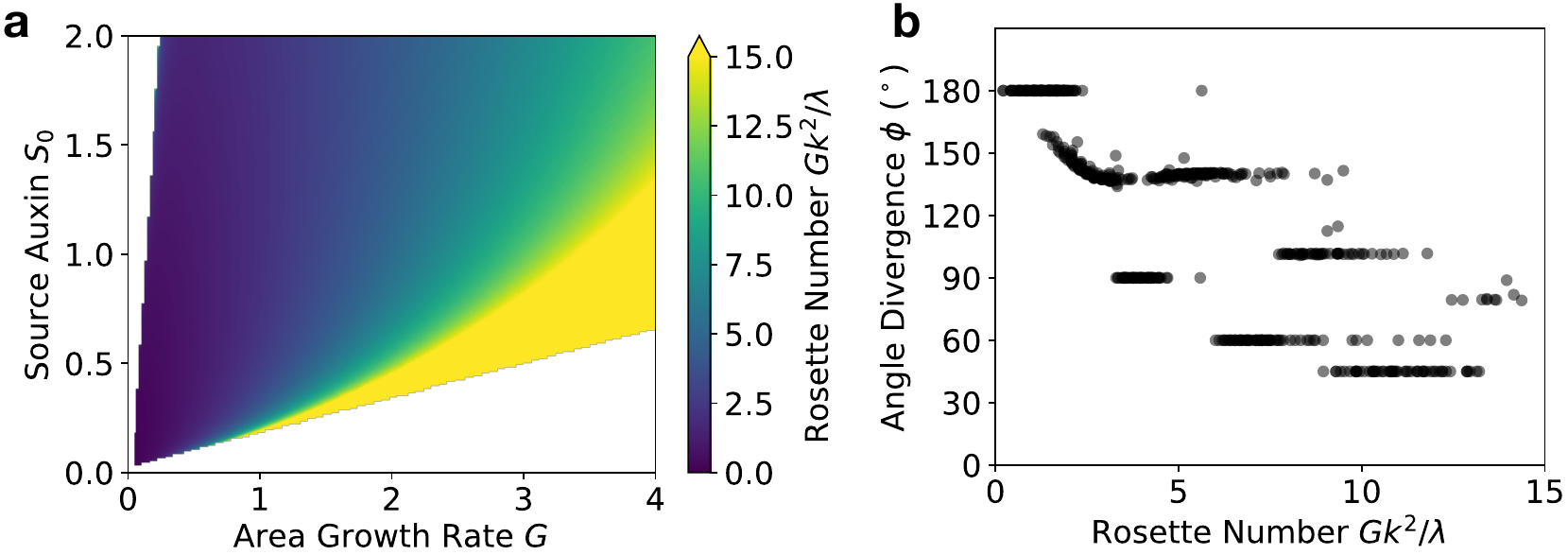}
\caption{Linear stability analysis can predict pattern type. (a) Rosette number $\frac{G k^2}{\lambda}$ against auxin source and growth rates. (b) The divergence against the Rosette number, for all values of $S_0$ and $G$ in Fig.~3. Points are transparent so darker regions show more overlapping points.}
\label{fig:4}
\end{figure}

% Talk about how it varies with S and G and the bounadaries
Calculating the Rosette number for each pair of auxin flux $S_0$ and area growth rate $G$, we observe a similar trend to the divergence, where increasing the area growth rate increases the Rosette number, while increasing the source decreases it (Fig.~\ref{fig:4}a). Notably, lines of constant Rosette number follow a similar upwards curve to the boundaries of different pattern types (Fig.~\ref{fig:4}b). Additionally, there are two regions which are stable against pattern formation, and so have an undefined Rosette number. As the area growth rate $G$ increases, the Rosette number increases and eventually diverges, suggesting that the pattern type would continue to increase up until the border of stability. From simulations, we observe that the distance at which peaks form increases with $G$, and so larger and larger simulation areas would be needed to capture this trend. In contrast, for low $G$ or high auxin source there is also a region without pattern forming, since saturation of auxin transport eventually stabilises when concentrations are very high.

Comparing our calculated Rosestte number against the divergence, we observe a fairly strong dependence between the two (Fig.~\ref{fig:4}b). Across all $S_0$ and $G$, increasing the Rosette number changes the divergence, and eventually the pattern type. While this is not a perfect fit, as the actual pattern observed doesn't fully depend on the Rosette number, such a simplified model considering only linear stability works remarkably well for most cases.

\section{Discussion}

As the molecular mechanisms which control phyllotaxis remains unclear, we demonstrate how a more general class of model, the Keller-Segel model, can reproduce various patterns observed in nature. Similar to the PIN1 models of phyllotaxis~\cite{heisler2010alignment, jonsson2006auxin, pennybacker2013phyllotaxis}, the Keller-Segel model is a self-organised patterning that has polarised transport of auxin, but the mechanism is through a signalling intermediate. There could be many mechanisms which produce this kind of instability leading to localised high concentrations of auxin, with mechanical interactions with polarisation being one of many. Thus, this model is similar but more general in that it demonstrates that the instability leading to auxin aggregation does not require mechanical interactions. Expanding upon previous work, our model also demonstrates how Fibonacci patterns may spontaneously form, without needing a pattern preseeded on the boundary~\cite{pennybacker2013phyllotaxis}, and shows how other patterns may be obtained depending on the growth rate and source of auxin.

By modelling the shoot apical meristem as a disc with a source of auxin at the center and radial advection due to growth, we show how the observed pattern depends on both that growth rate and the source at the center. In addition to forming alternating patterns which then twist into Fibonacci spirals, as in the Douady and Couder experiments~\cite{douady1992phyllotaxis}, this system can also form whorled patterns, where two or more or more peaks form at the same distance from the center. In contrast to previous mechanical and biochemical models by Newell and collaborators~\cite{newell2008phyllotaxis}, our model spontaneously forms phyllotactic patterns from inside-out, rather than initiating the system with a Fibonnaci spiral on the outside to form a pattern outside-in. Finally, we develop a simple value, the Rosette number, formed using linear stability analysis which can explain most of the patterns observed. This model is theoretical and uses an unidentified transport signal $c$, it acts to direct transport of auxin and could, for example, represent some hormone which directs PIN1 polarisation.

One aspect of the model that is not investigated in detail here is the role of the growth profile and the geometry of the surface. Within our model we have modelled growth as occurring only at the center, which leads to zero dilution in the auxin profile. In reality, the plant cells will continue growing away from the center, as in the exponential growth rate used by Doaudy and Couder~\cite{douady1992phyllotaxis}, which would dilute the auxin profile with distance. This would both change the growth rate of any instabilities and cause them to be potentially transient, meaning waves which do not grow fast enough are diluted and become stable, potentially limiting the range of pattern types observed. Similarly, the geometry of the system can also control the dilution rate. Here we have assumed a flat surface, which may be appropriate for a sunflower, but other plants such as tree branches or cacti may be curved at the tip and cylindrical around the stem, which would initially dilute any auxin before preserving concentrations around the stem. In this case, nonlinear effects would be important for maintaining the arrangement of primordia. Moreover, these would grow and potentially act as their own source of auxin, which may change the interaction strength with nearby peaks.

%Preliminary simulations using a constant speed growth profile, or a paraboloid surface, show that they can still produce whorled and Fibonacci spirals, with the pattern type increase with the growth rate, suggesting that the patterns obtained are a universal feature, with geometry and growth controlling the transition points. [need supplementary figure? These are the old simulations so they can be done quickly] 

\section{Acknowledgements}

I thank Boris Shraiman for inspiring this project, and his valuable feedback and discussions.

\section{Appendix}

\subsection{Linear Stability Analysis}

At steady state, we have a constant flux
\begin{equation}
    \nabla \cdot (a \mathbf{v}) = 0.
\end{equation}
Solving for $a$ we get
\begin{equation}
    a(r) = \bar{a} = \frac{\pi r_s^2 S_0}{2 \pi G}.
\end{equation}
Since the disc has a uniform auxin concentration, and the advection conserves concentrations, we perform a linear stability analysis to a system of uniform auxin concentration $a_0$ and transport signal concentration $\beta a_0$:
\begin{equation}
    \dot{a} = \nabla^2 a - \alpha \nabla \cdot \left(a \nabla \frac{c}{c + c_0}\right)
\end{equation}
\begin{equation}
    \dot{c}= D\nabla^2 c + \beta a - c
\end{equation}
We apply a small perturbation about this such that $a = a_0 + \tilde{a}$ and $c = \beta a_0 + \tilde{c}$. These evolve, to leading order, as
\begin{equation}
    \dot{\tilde{a}} = \nabla^2 \tilde{a} - \alpha \frac{a_0 c_0}{(\beta a_0 + c_0)^2} \nabla^2 \tilde{c}
\end{equation}
and 
\begin{equation}
    \dot{\tilde{c}} = D \nabla^2 \tilde{c} + \beta \tilde{a} - \tilde{c}.
\end{equation}
Working in Fourier space we have
\begin{equation}
    \partial_t \hat{a} = -k^2 \hat{a} + k^2 \frac{\alpha a_0 c_0}{(\beta a_0 + c_0)^2} \hat{c}
\end{equation}
\begin{equation}
    \partial_t \hat{c} = \beta \hat{a} - (D k^2 + 1) \hat{c}.
\end{equation}

This system of equations has eigenvalues
\begin{equation}
\begin{split}
\lambda_\pm =& -\frac{1}{2} ((1 + D)k^2 + 1) \\
&\pm \frac{1}{2} \sqrt{((D+1)k^2 + 1)^2 + 4 k^2 (\alpha' - (Dk^2 + 1))}.
\end{split}
\end{equation}
where $\alpha' = \frac{\alpha \beta a_0 c_0}{(\beta a_0 + c_0)^2}$ is the effective transport rate due to saturation. The eigenvalue for a given $k$ will be positive when $\frac{\alpha \beta a_0 c_0}{(\beta a_0 + c_0)^2} > (D k^2 + 1)$ and so we always have an instability when $\alpha \beta a_0 c_0 > (\beta a_0 + c_0)^2$. From this, one can find the fastest growing wave length, with growth rate $\lambda^*$ and corresponding wave number $k^*$, which are found to be
%Credit to wolfram alpha: "maximise  -((1 + d)x^2 +1) + (((1 + d) * x^2 + 1)^2 + 4* x^2 * (a - d * x^2 - 1))^0.5 wrt x"
% \begin{equation}
%     \lambda^* = \frac{1}{(D-1)^2}\left(\alpha'D + \alpha' + D - 1 -2\sqrt{\alpha' D(\alpha' + D - 1)}\right)
% \end{equation}
\begin{equation}
    \lambda^* = \frac{1}{(D-1)^2}\left(\sqrt{\alpha' D} - \sqrt{\alpha' + D - 1}\right)^2
\end{equation}
and
\begin{equation}
    k^* = \sqrt{\frac{\sqrt{D(D+1)^2 (a^2 + a(D-1))} + D(1 - D - 2a)}{D (D-1)^2}}
\end{equation}

\begin{ruledtabular}
\begin{tabular}{ccc}
 Parameter & Symbol & Default Value\\
\hline
Area growth rate & $G$ & 1\\
Auxin transport rate & $\alpha$ & $10$\\
Saturating $c$ & $c_0$ & 5\\
$c$ Production Rate & $\beta$ & 2\\
Auxin flux rate & $S_0$ & 1\\
Source Timescale & $\tau$ & 100\\
Source radius & $r_s$ & 2\\
Disc Radius & $r_d$ & 50\\
\end{tabular}
\end{ruledtabular}

\bibliography{refs}% Produces the bibliography via BibTeX.

\end{document}